\documentclass[prb,twocolumn,showpacs,amsmath,amssymb,superscriptaddress]{revtex4-2}
\usepackage{mathtools}
\usepackage{bm} 
\usepackage{dcolumn} 
\usepackage{graphicx} 
\usepackage{amsmath} 
\usepackage{amssymb} 
\usepackage{amsfonts} 
\usepackage{placeins}

\newcommand{\mdos}[1]{$m_\mathrm{DOS}$}

\begin{document}
\title{Screening, Friedel oscillations, RKKY interaction, and Drude transport in anisotropic two-dimensional systems}

\author{Seongjin Ahn}
\affiliation{Condensed Matter Theory Center and Joint Quantum Institute, Department of Physics, University of Maryland, College Park, Maryland 20742-4111, USA}
\author{S. Das Sarma}
\affiliation{Condensed Matter Theory Center and Joint Quantum Institute, Department of Physics, University of Maryland, College Park, Maryland 20742-4111, USA}

\date{\today}

\begin{abstract}
We investigate the effect of the mass anisotropy on Friedel oscillations, Ruderman-Kittel-Kasuya-Yosida (RKKY) interaction, screening properties, and Boltzmann transport in two dimensional (2D) metallic and doped semiconductor systems. We calculate the static polarizability and the dielectric function within the random phase approximation with the mass anisotropy fully taken into account without making any effective isotropic approximation in the theory. We find that carrier screening exhibits an isotropic behavior for small momenta despite the anisotropy of the system, and becomes strongly anisotropic above a certain threshold momentum. Such an anisotropy of screening leads to anisotropic Friedel oscillations, and an anisotropic RKKY interaction characterized by a periodicity dependent on the direction between the localized magnetic moments. We also explore the disorder limited dc transport properties in the presence of mass anisotropy based on the Boltzmann transport theory. Interestingly, we find that the anisotropy ratio of the short-range disorder limited resistivity along the heavy- and light-mass directions is always the same as the mass anisotropy ratio whereas for the long-range disorder limited resistivity the anisotropy ratio is the same as the mass ratio only in the low density limit, and saturates to the square root of the mass ratio in the high density limit. Our theoretical work should apply to many existing and to-be-discovered anisotropic 2D systems. 
\end{abstract}

\maketitle

\section{Introduction} \label{sec:Introduction}
There has been considerable interest, going back to the early 1970s, in two-dimensional electron gas (2DEG) systems due to their importance both in pure theoretical studies and technological applications \cite{Ando1982}. The integer quantum Hall effect was originally observed in 2D Si-SiO$_2$ based inversion layers \cite{Klitzing1980} and then in many 2D systems over the last 40 years. Subsequently, the fractional quantum Hall effect was seen in 2D GaAs-AlGaAs systems, and later in many other systems \cite{Tsui1982}. The 2DEG is typically formed at the interface between bulk materials \cite{Ando1982}, where electrons are confined to a potential that restricts and quantizes their motion along one direction, allowing only two degrees of freedom along the 2D interface plane. More recently, graphene and related materials provide examples of intrinsic 2D materials made of electrons confined in one atomic monolayer. 
Very recently, researchers successfully fabricated atomically thin 2D materials (e.g., black phosphorus \cite{Gusmao2017, Deng2014, Wang2015}, transition metal dichalcogenides \cite{Manzeli2017}, etc.) with the help of advanced techniques that enable to exfoliate a single layer of atoms from the bulk layered counterpart \cite{Novoselov2005}.

The most common theoretical approach to understanding physics in 2D materials is to start from the ideal isotropic 2DEG model where the energy dispersion is given by $\varepsilon_{\bm k}=\frac{\hbar^2k^2}{2m}$. The effective mass $m$ is determined by an isotropic averaged mass such as density-of-states mass or conductivity mass \cite{Beni1978, Luttinger1955, Brinkman1972, Combescot2001}, depending on the carrier property being studied. Here, the density of states mass, $m_\mathrm{DOS}$, and the conductivity (or optical) mass, $m_\mathrm{CON}$, are defined by: $m_\mathrm{DOS} = (m_x m_y)^{1/2}$ and $m_\mathrm{CON}= 2 m_x m_y/(m_x + m_y)$, where $m_{x,y}$ are the anisotropic effective masses along the 2D cartesian axes. We mention that $m_\mathrm{DOS}$ typically appears in thermodynamic quantities such as the specific heat whereas $m_\mathrm{CON}$ appears in transport and optical properties such as the conductivity. Despite its simplicity and neglect of the anisotropy, the isotropic 2D model has been reasonably successful in capturing many physical properties of anisotropic 2D systems \cite{Ando1982, Brinkman1972, Brinkman1972a, Ting1975, Lee1975, Zheng1996, Priour2005}. Recent work, however, has demonstrated that such a neglect of mass anisotropy could lead to the incorrect suppression of key anisotropic features that have no corresponding isotropic analog \cite{Ahn2020, Ahn2021, Ahn2021a,  Roldan2006}. In addition, there have been several reports of mass anisotropy in new emerging 2D materials due to the low in-plane symmetry, with the observation of rich anisotropic physics absent in isotropic materials \cite{ Yang2018, Chenet2015, Wang2015, Huang2016,  Li2017, Qiu2016}. Thus it is imperative to develop a general electric theory incorporating the mass anisotropy to correctly describe physical properties of 2D electronic materials.

In this paper, we discuss the effect of the mass anisotropy on screening, Friedel oscillations, RKKY interaction, and dc Drude transport by taking the explicit effective mass anisotropy into account using the anisotropic 2DEG model described by
\begin{equation}
    \varepsilon_{\bm k}=\frac{k_x^2}{2m_\mathrm{H}}+\frac{k_y^2}{2m_\mathrm{L}},
\end{equation}
where $m_\mathrm{H}$ and $m_\mathrm{L}$ denote the heavy and light masses, respectively, taken without loss of generality to be along the cartesian coordinates $x$ and $y$ in the 2D plane of confinement. We provide detailed analytical and numerical analysis of anisotropic behaviors of these properties, comparing them with the extensively used and well-known isotropic results where anisotropic masses are averaged into a single effective isotropic effective mass. We also discuss the validity of the isotropic approximation by exploring several regimes of electron density where the isotropic approximation works (or fails). We find that, contrary to prevalent expectations, the isotropic approximation often leads to incorrect and misleading results.


This paper is organized as follows. In Sec.~\ref{sec:static_screening}, we discuss screening properties of anisotropic 2DEG, presenting analytic results for the polarizability and the dielectric function, highlighting their anisotropic features. We also discuss the Friedel oscillations associated with 2$k_\mathrm{F}$-screening in Sec.~\ref{sec:static_screening}. From the obtained screening results in Sec.~\ref{sec:static_screening}, we calculate the RKKY interaction between localized magnetic moments \cite{Yosida1957,Kasuya1956,Ruderman1954} in Sec.~\ref{sec:RKKY} and analyze anisotropic effects on its oscillatory behavior. Section \ref{sec:transport} presents the calculated disorder scattering induced carrier resistivity results obtained using the Boltzmann transport equation with the anisotropic electron structure fully taken into account. We consider both short-range defect and long-range Coulomb disorders, and provide detailed analytical and numerical analysis of the resistivity behaviors for both types of scatterer. Section~\ref{sec:conclusion} contains a summary and conclusions. Our theory is entirely restricted to the zero temperature situation although a finite temperature generalization is straightforward and cumbersome, by using finite temperature Fermi distribution function and chemical potential everywhere.

\section{anisotropic screening} \label{sec:static_screening}
The 2D static screening function in the random phase approximation (RPA) is given by 
\begin{equation}
\varepsilon(\bm q)=1-v_c(\bm q)\Pi_0(\bm q)
\label{eq:dielectric_function}
\end{equation}
where $v_c(q)=\frac{2\pi e^2}{q}$ is the 2D Coulomb interaction (note that the Coulomb interaction itself is isotropic even in the presence of mass anisotropy), $q=\sqrt{q_x^2+q_y^2}$ and $\Pi_0(\bm q)$ is the noninteracting irreducible static polarizability given by 
\begin{equation}
    \Pi_0(\bm q) = \int \frac{d^2k}{(2\pi)^2} \frac{n_\mathrm{F}(\xi_{\bm k})-n_\mathrm{F}(\xi_{\bm k+\bm q} )}{\varepsilon_{\bm k}-\varepsilon_{\bm k+\bm q}}.
\end{equation}
where $n_\mathrm{F}(\xi_{\bm k})$ is the Fermi-Dirac distribution function, $\xi_{\bm k}=\varepsilon_{\bm k}-\mu$, and $\mu$ is the chemical potential. The polarizability for an anisotropic 2D electron gas can be obtained using the existing result for the isotropic polarizability \cite{Stern1967} by rescaling $m_e\rightarrow m_\mathrm{DOS}$, $q_x\rightarrow\sqrt{\frac{m_\mathrm{DOS}}{m_\mathrm{H}}} q_x$ and $q_x\rightarrow\sqrt{\frac{m_\mathrm{DOS}}{m_\mathrm{L}}} q_y$, where $m_\mathrm{e}$ is the electron mass. We obtain the anisotropic static polarizability function to be
\begin{align}     \label{eq:polar}
    \Pi_0(\bm q)=-D(E_\mathrm{F})
		\left[1 - \Theta(q_\mathrm{iso}-2k_F^\mathrm{DOS})\frac{\sqrt{q_\mathrm{iso}^2- \left(2k_F^\mathrm{DOS}\right)^2 }}{q_\mathrm{iso}} \right],
\end{align}
where $D(E_\mathrm{F})=\frac{m_\mathrm{DOS}}{\pi\hbar}$ is the 2D density-of-states and $q_\mathrm{iso}$ is defined as satisfying
\begin{equation} \label{eq:qiso}
	\frac{q_\mathrm{iso}^2}{2m_\mathrm{DOS}}=\frac{q_x^2}{2m_\mathrm{H}}+\frac{q_y^2}{2m_\mathrm{L}}.
\end{equation}

\begin{figure}[htb]
  \centering
  \includegraphics[width=0.95\columnwidth]{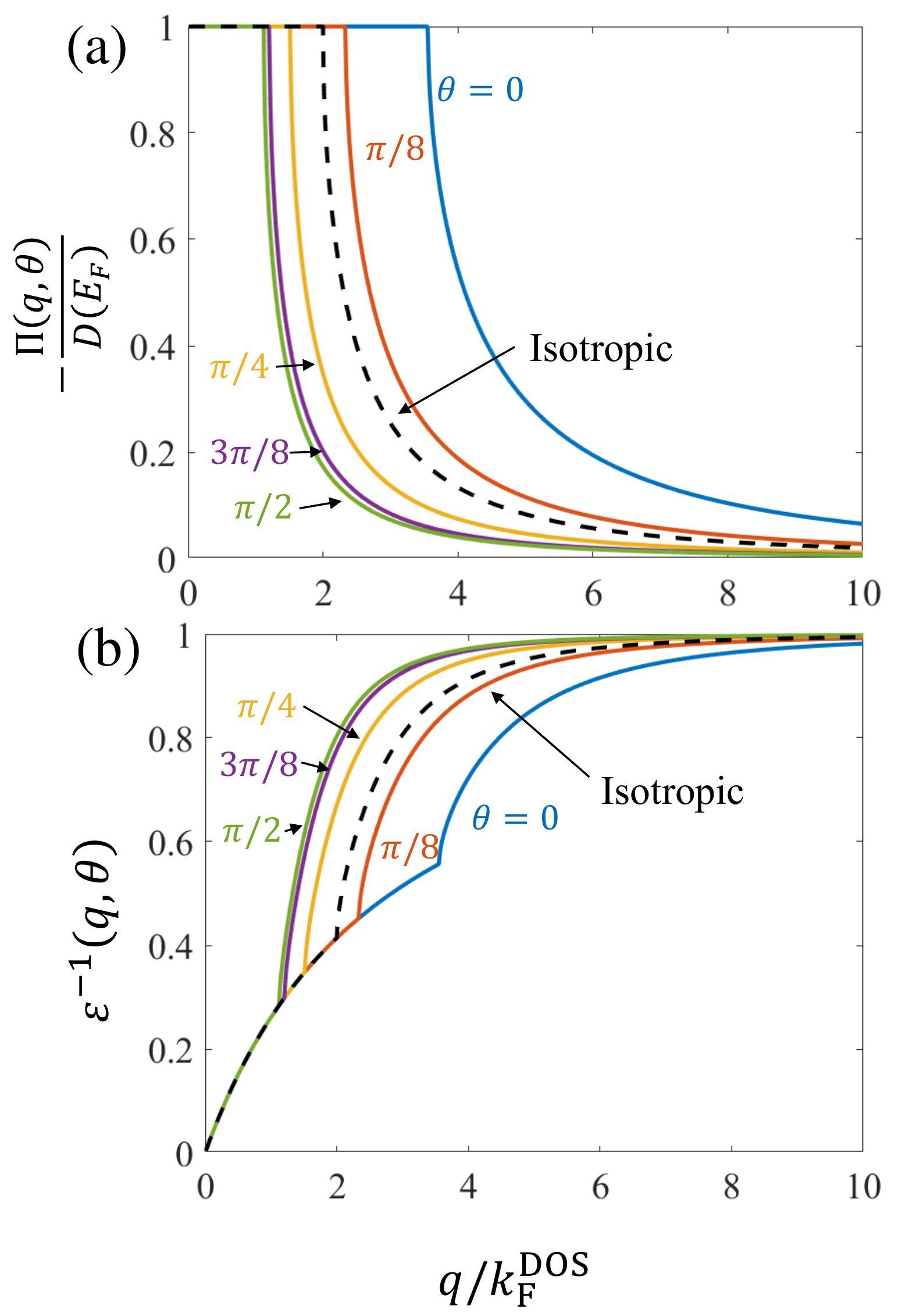}
  \caption{(a) the polarizability and (b) the dielectric function plotted along several different directions $\theta=0$, $\pi/8$, $\pi/4$, $3\pi/8$ and $\pi/2$ where $\theta$ is the angle from the heavy mass axis. Here we set $r_s=2.0$ and $m_\mathrm{H}/m_\mathrm{L}=10$. }\label{fig:polarizability}
\end{figure}
	
Figure \ref{fig:polarizability}(a) shows the polarizability function plotted along several different directions. All the curves show the typical 2D polarizability behavior  \cite{Stern1967}, being a constant up to a certain critical momentum $q_c(\theta)$ at which the polarizability suddenly drops, resulting in a kink structure characterized by the discontinuity in its first derivative, and falls off rapidly ($\sim1/q^2$) with $q$ increasing further. Here $\theta$ is the angle from the heavy-mass direction (i.e., $x$-axis) defined through $\theta=\tan^{-1}(q_y/q_x)$. The constancy of the low momentum screening function in Fig.~\ref{fig:polarizability} up to a finite momentum is directly connected to the constant energy-independent form of the 2D density of states. At large momenta, the anisotropy is strongly suppressed because of the decay of the short distance (i.e. large $q$) screening. $q_c(\theta)$ can be obtained analytically using Eq.~(\ref{eq:qiso}) with $q_\mathrm{iso}$ substituted by $2k_\mathrm{F}^\mathrm{DOS}$: 

\begin{equation}
	q_c(\theta)=2k_\mathrm{F}(\theta)
	\label{eq:singularity_point}
\end{equation}
where 
\begin{equation}
	k_\mathrm{F}(\theta) = \frac{k_\mathrm{F}^\mathrm{DOS}}
	{ \sqrt { 
		\frac{m_\mathrm{DOS}}{m_\mathrm{H}}\cos^2\theta +
		\frac{m_\mathrm{DOS}}{m_\mathrm{L}}\sin^2\theta}},
\label{eq:Fermi_wavevector}
\end{equation}
For small $q$, more precisely $q<q_c(\pi/2)=\sqrt{\frac{m_\mathrm{L}}{m_\mathrm{DOS}}}k_\mathrm{F}^\mathrm{DOS}$, the polarizability is constant and isotropic with $\Pi_0(\bm q)=-\frac{m}{\pi\hbar}$ despite the mass anisotropy of the system. Note that for the isotropic system with $m_\mathrm{H}=m_\mathrm{L}$, the screening function is constant up to 2$k_\mathrm{F}$, where $k_\mathrm{F}$ is the isotropic Fermi momentum. Thus, $k_\mathrm{F}(\theta)$ is simply the effective anisotropic Fermi momentum defined by Eq.~(\ref{eq:Fermi_wavevector}).

In Fig.~\ref{fig:polarizability}(b), we present the corresponding dielectric function plotted along several different directions. Note that the screening is isotropic in the long-wavelength limit due to the isotropic polarizability as discussed above. Using Eq.~(\ref{eq:polar}), it is easy to see that the dielectric function in the long wavelength limit is given by
\begin{equation}
	\varepsilon(\bm q)=1 + \frac{q_\mathrm{TF}}{q}
	\label{eq:Thomas_Fermi_dielectric}
\end{equation} 
where $q_\mathrm{TF}=2m_\mathrm{DOS}e^2/\hbar^2$ is the Thomas Fermi wavevector. Note that Eq.~(\ref{eq:Thomas_Fermi_dielectric}) is indeed identical to the long-wavelength dielectric function of a 2D isotropic electron gas with only the effective mass $m$ replaced by the density of states mass $m_\mathrm{DOS}$. This shows that for sufficiently small $q$ the isotropic approximation employing the density-of-states mass works well in describing the screening properties of the anisotropic system. 
This may be the underlying reason for the widespread practice of ignoring anisotropy effects in the screening properties of anisotropic metals and doped semiconductors since the long wavelength Thomas-Fermi screening is the standard approximation used extensively.  We point out that the long wavelength Thomas-Fermi screening is strictly proportional to the density of states of the system which for the anisotropic system is indeed defined by the density-of-states effective mass, explaining why screening becomes isotropic in the small momentum limit. At large momentum, obviously the isotropic approximation fails as described in this work.

 It is worth noting that the finite wavevector screening is stronger along the the heavy mass direction ($\theta=0$) than that along the light-mass direction ($\theta=\pi/2$). This can be understood as follows: for an anisotropic electron gas, we can define two different Wigner-Seitz radius parameter $r_s^\mathrm{H}$ and $r_s^\mathrm{L}$, which represent the effective strength of the Coulomb interaction along the heavy mass and the light mass directions, respectively. Since $r_s^\mathrm{H}>r_s^\mathrm{L}$, which can be easily seen by noting that $r_s \sim m$, the screening along the heavy direction should be stronger in general than that along the light-mass direction. An equivalent physical way of explaining this is that screening being proportional to the effective mass in the long wavelength limit (through its proportionality on the density of states), it is stronger (weaker) in the direction of heavier (lighter) mass.

Before concluding this section, it is worth investigating the consequence of the directional-dependent kink behavior of the polarizability on the real space properties of anisotropic screening behavior.  The 2$k_\mathrm{F}$-kink in the screening function arises directly from the discontinuity in the Fermi function at $k_\mathrm{F}$, and is closely related to the Kohn anomaly \cite{Kohn1959}. Anisotropy obviously maintains the sharpness of the Fermi surface, and hence the 2$k_\mathrm{F}$ kink is preserved anisotropically [see Eq.~(\ref{eq:kp})] in the polarizability.
One of the well-known consequences of the kink is electron density oscillation near a charged impurity, known as Friedel Oscillation, arising from the 2$k_\mathrm{F}$ kink in screening. The effective screened potential at large distances ($rk^\mathrm{DOS}_\mathrm{F}\gg1$) from a charged impurity in an anisotropic electron gas is calculated to be
\begin{align}
	\phi(\bm r)&=\int \frac{d^2 q}{(2\pi)^2} \frac{v_c(\bm q)}{\varepsilon(\bm q)}e^{i \bm q \cdot \bm r}\notag\\
                        &\sim \frac{4q_\mathrm{TF}(k^\mathrm{DOS}_\mathrm{F})^2}{(2k^\mathrm{DOS}_\mathrm{F} + q_\mathrm{TF})^2} 							  \frac{\sin[2k_\mathrm{P}(\theta)r]}{[2k_\mathrm{P}(\theta)r]^2} ,
\label{eq:Friedel_Oscillation}
\end{align}
where 
\begin{equation}
	k_\mathrm{P}(\theta)=k_\mathrm{F}^\mathrm{DOS}\sqrt{ \frac{m_\mathrm{H}}{m_\mathrm{DOS}}\cos^2\theta
						 +\frac{m_\mathrm{L}}{m_\mathrm{DOS}}\sin^2\theta},
\label{eq:kp}
\end{equation}
and $r=\sqrt{x^2+y^2}$ and $\theta=\tan^{-1}(y/x)$. Note that due to the anisotropy of the polarizability, the oscillation term in Eq.~(\ref{eq:Friedel_Oscillation}) is highly anisotropic varying as a function of $\theta$.
The oscillation along the heavy-mass and light-mass direction has a periodicity $\pi/k^\mathrm{H}_\mathrm{F}$ and $\pi/k^\mathrm{L}_\mathrm{F}$, respectively, where $k^\mathrm{H}_\mathrm{F}$ ($k^\mathrm{L}_\mathrm{F}$) is the magnitude of the Fermi-wavevector along the heavy-mass, i.e., $x$-axis, (light-mass, i.e., $y$-axis) direction. The period of oscillations along an arbitrary direction off the symmetry axis lies in between $\pi/k^\mathrm{H}_\mathrm{F}$ and $\pi/k^\mathrm{L}_\mathrm{F}$. Such a direction-dependent Friedel oscillation is a clear prediction of our theory which should manifest itself in 2D anisotropic systems, e.g., Kohn anomaly should manifest strong angular dependence in the 2D plane.

The singularity in the polarizability also plays an important role in the interaction between localized magnetic moments mediated by the itinerant electrons, the so called ``RKKY interaction" \cite{Yosida1957,Kasuya1956,Ruderman1954}. In the next section, we briefly introduce the formalism for the RKKY interaction showing its relation to the polarizability, and discuss anisotropic features of the RKKY interaction arising from mass anisotropy.


\section{RKKY interaction} \label{sec:RKKY}

\begin{figure}[htb]
  \centering
  \includegraphics[width=0.95\columnwidth]{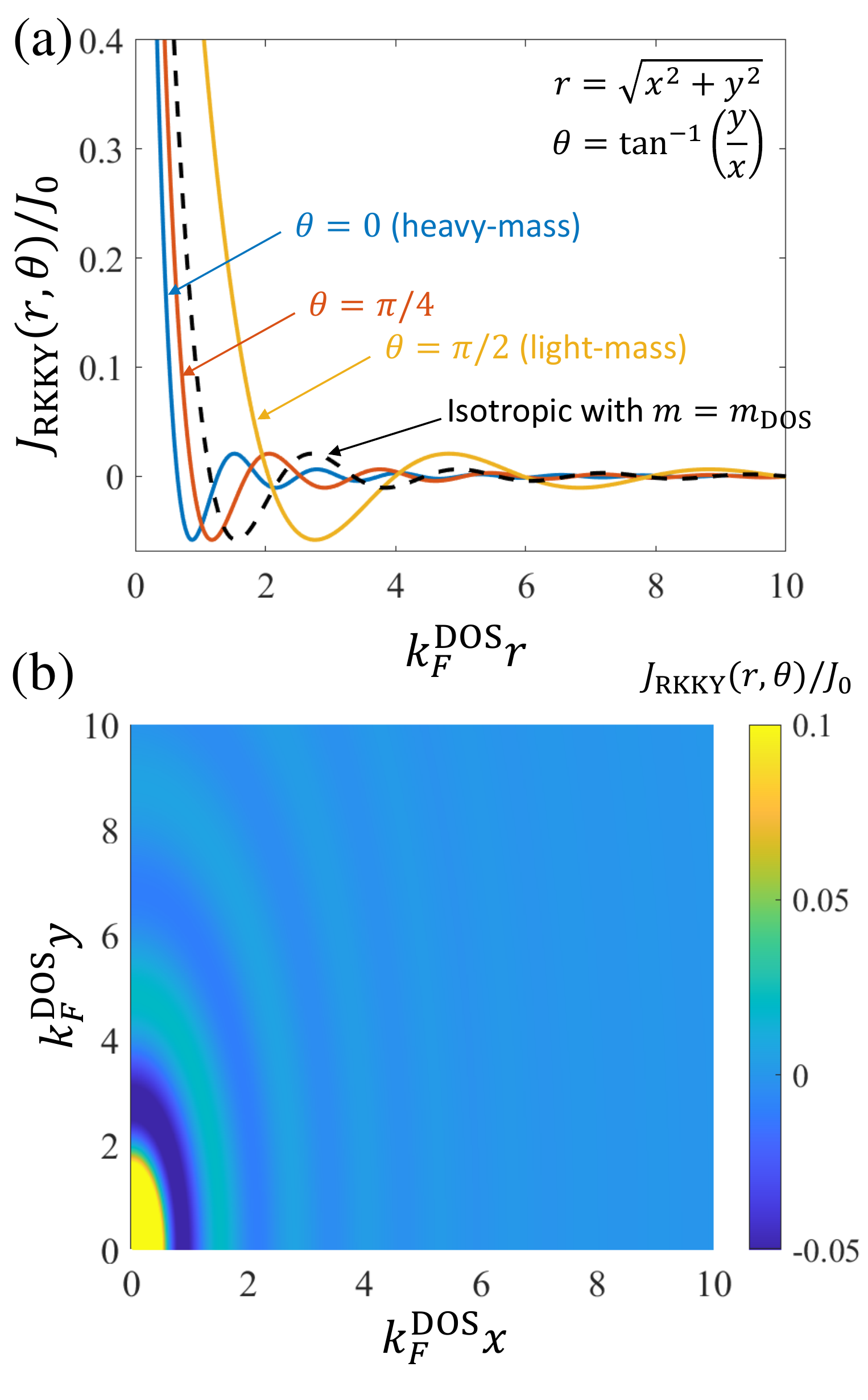}
  \caption{Plots of $J(r,\theta)$ for several different directions $\theta=0$, $\pi/4$, and $\pi/2$ along with the isotropic one obtained using $m_\mathrm{DOS}$ (black-dashed line). The bottom figure shows two dimensional plot of $J(r,\theta)$, highlighting the anisotropy of RKKY interaction.  Here $J_0=-\frac{J_\mathrm{ex}^2}{4} D(E_\mathrm{F}) (k_\mathrm{F}^\mathrm{DOS})^2. $}\label{fig:RKKY}
\end{figure}

The localized spin moment interacting with an itinerant electron via the exchange interaction is described by \cite{Ruderman1954, Kasuya1956,Yosida1957, Patrone2012} 
\begin{equation}
H_\mathrm{ex}=-J_\mathrm{ex} \sum_{i=1,2} \bm S_i \cdot \bm s (\bm R_i),
\end{equation}
where $J_\mathrm{ex}$ is the exchange coupling constant, $\bm S_i$ is the magnetic impurity moment located at $\bm R_i$ and $\bm s (\bm r)=\frac{\hbar}{2} \sum_i \delta(\bm r- \bm r_i)\sigma_i$ is the spin density for an electron located at $\bm r$. Treating $H_\mathrm{ex}$ as a perturbation and expanding it up to the second order, we obtain 
\begin{equation}
H_\mathrm{RKKY}=\sum_{i,j}J_\mathrm{RKKY}(\bm r)\bm S_i \cdot \bm S_j
\label{eq:HRKKY}
\end{equation}
where $\bm r= \bm R_i - \bm R_j $ and 
\begin{equation}
	J_\mathrm{RKKY}(\bm r)=\frac{J_\mathrm{ex}^2}{4}\Pi(\bm  r).
\label{eq:JRKKY}
\end{equation}
Eq.~(\ref{eq:HRKKY}) and Eq.~(\ref{eq:JRKKY}) show that the exchange-mediated interaction between two magnetic impurities is proportional to $\Pi(\bm  r)$, which is the real space Fourier transform of the static polarizability in the momentum space given by  Eq.~(\ref{eq:polar}), i.e.,

\begin{align}
	\Pi(\bm r)&=\int \frac{d^2 q}{(2\pi)^2}\Pi(\bm q) e^{i \bm q \cdot \bm r}\notag\\
			   &= -\frac{2(k_\mathrm{F}^\mathrm{DOS})^2}{m_\mathrm{DOS} \hbar}\times \notag\\
                      &\left\{  
                      J_0[k_\mathrm{p}(\theta)r]
                      N_0[k_\mathrm{p}(\theta)r] + J_1[k_\mathrm{p}(\theta)r]
                      N_1[k_\mathrm{p}(\theta)r]
                      \right\},
\label{eq:real_polar}
\end{align}
which is similar to the isotropic counterpart \cite{Aristov1997} with the isotropic Fermi wavevector replaced by $k_\mathrm{p}(\theta)$.
Here $r=\sqrt{x^2+y^2 }$, and $J_n(x)$ and $N_n(x)$ are Bessel functions of the first and second kind, respectively.

Figure.~\ref{fig:RKKY}(a) presents plots of RKKY interaction $J_\mathrm{RKKY}(\bm r)$ along various directions. For small $r<1/k_\mathrm{F}^\mathrm{DOS}$, the RKKY interaction is stronger along the heavy-mass direction than along the light-mass direction, and starts showing an oscillatory behavior as $r$ increases. The oscillation period along the light mass axis is much larger than that along the heavy-mass direction, similar to the Friedel oscillation discussed in the previous section. It is also worth noting that the RKKY interaction decays much more rapidly along the heavy-mass direction than along the light-mass direction. These results show that the RKKY interaction is anisotropic in the presence of mass anisotropy, strongly deviating from the isotropic result obtained using the isotropic approximation (black-dashed line). This clear prediction of our theory should be directly observable in 2D anisotropic metals and doped systems.

To understand the anisotropic features of the RKKY interaction, here we provide the asymptotic form of the real space polarizability  [Eq.~(\ref{eq:real_polar})]. In the small distance limit ($k_\mathrm{F}r\ll1$), Eq.~(\ref{eq:real_polar}) is written as
\begin{equation}
    \Pi(\bm r) \approx -\frac{2(k_\mathrm{F}^\mathrm{DOS})^2}{m_\mathrm{DOS} \hbar}
    \frac{2\gamma-1 - 2\ln[k_\mathrm{p}(\theta)r/2]}{\pi},
    \label{eq:asymptotic_RKKY_short}
\end{equation}
where $\gamma$ is the Euler constant. This shows that the RKKY interaction diverges logarithmically as $r\rightarrow0$ with its diverging rate along the heavy mass direction ($\sim \ln rk^\mathrm{H}_\mathrm{F} $) being much faster than that along the light-mass direction ($\sim \ln rk^\mathrm{L}_\mathrm{F} $). Also note that since $k_\mathrm{p}(\theta)$ increases with increasing $\theta$, one can easily see from Eq.~(\ref{eq:asymptotic_RKKY_short}) that the RKKY interaction becomes stronger as one moves from the heavy-mass axis (i.e., $x$-axis) to the light mass axis (i.e., $y$-axis).
In the long distance limit, i.e., $k_\mathrm{F}r\gg1$: 
\begin{equation}
	\Pi(\bm r)\approx -\frac{8(k_\mathrm{F}^\mathrm{DOS})^2}{m_\mathrm{DOS} \hbar\pi}      \frac{\sin[2k_\mathrm{P}(\theta)r]}{[2k_\mathrm{P}(\theta)r]^2}.
\label{eq:asymptotic_RKKY}
\end{equation}
Note that the period of oscillation is given not by $\pi/k_\mathrm{F}(\theta)$ but by $\pi/k_\mathrm{p}(\theta)$, implying that unlike the isotropic case the RKKY oscillation period for an anisotropic system is not necessarily determined by the Fermi wavevector. 
Along the symmetry axis, however, the period of oscillation is associated with the Fermi wavevector, given by $\pi/k^\mathrm{H}_\mathrm{F}$ and $\pi/k^\mathrm{L}_\mathrm{F}$ along the heavy- and the light-mass directions, respectively. 
Also note that whereas the RKKY interaction decays with the same power law ($\sim1/r^2$) along all directions, the decay rate depends on the direction, being much faster along the heavy-mass direction with $\sim1/(2k_\mathrm{F}^\mathrm{H}r)^2$ than along the light-mass direction $\sim1/(2k_\mathrm{F}^\mathrm{L}r)^2$ as determined by the 2D anisotropy of the effective mass.

Figure~\ref{fig:RKKY}(b) shows the two-dimensional plot of the RKKY interaction in the $x$-$y$ plane. This figure highlights that the RKKY interaction is highly anisotropic, showing that the period of the oscillation varies as a function of the direction between the localized magnetic moments.

\section{Transport} \label{sec:transport}
In this section, we investigate the dc transport properties of an anisotropic electron gas, arising from impurity scattering, using the Boltzmann-transport equation within the relaxation time approximation. For an isotropic system, the Boltzman equation is well known to provide the dc transport scattering rate to be \cite{Ashcroft1976}
\begin{equation}
\frac{1}{\tau^\mathrm{iso}_{\varepsilon_{\bm k} } }
=
\int
\frac{d^2k^\prime}{(2\pi)^2}
W_{\bm k \bm k^\prime}
\left(
1 -
\cos\theta_{\bm k \bm k^\prime}
\right)
\label{eq:isotropic_relaxation_time}
\end{equation}
where the disorder scattering matrix element is given by
\begin{equation}
W_{\bm k \bm k^\prime}
=
\frac{2\pi}{\hbar}
n_\mathrm{imp}
\left| V_{\bm k \bm k^\prime}
\right |^2
\delta(\varepsilon_{\bm k} - \varepsilon_{\bm k^\prime}).
\label{eq:matrix_element}
\end{equation}
and $\cos\theta_{\bm k \bm k^\prime}$ is the angle between $\bm k$ and $\bm k^\prime$.
Here $n_\mathrm{imp}$ denotes the impurity density and $\left| V_{\bm k \bm k^\prime}\right |$ is the impurity potential. One should be cautious when applying this equation since it has been shown that the equation can lead to inaccurate transport results in the presence of an anisotropic Fermi surface, and thus should instead use a modified version of the Boltzmann transport formalism in an integral equation form \cite{Vyborny2009}. Basically, the relaxation time approximation must be carried out incorporating the anisotropy explicitly, which is straightforward to do. For completeness, we briefly introduce the derivation of the anisotropic Boltzmann transport theory before we present our results. 

In this paper, we assume that the system is stationary and homogeneous so that spatial and temporal changes of the distribution after collisions are negligible, i.e., $f(\bm r + \bm v dt, \bm k + \bm Fdt, t+dt) = f(\bm r, \bm k + \bm Fdt, t)$ where $\bm v$ is the velocity and $\bm F$  is the Lorentz force acting on electron in an external uniform and static electric field. Then, the difference in the distribution over time $dt$ induced by scattering is given by $f(\bm r + \bm v dt, \bm k + \bm Fdt, t+dt) - f(\bm r, \bm k, t)= e\bm E \cdot \bm v dt \frac{ \partial f}{ \partial \varepsilon_{\bm k}}$, leading to the Boltzmann equation 
\begin{equation}
    e\bm E \cdot \bm v  \frac{ \partial f}{ \partial \varepsilon_{\bm k}} = \left(\frac{\partial f}{\partial t}\right)_\mathrm{coll}
    \label{eq:BE}
\end{equation}
where $\left(\frac{\partial f}{\partial t}\right)_\mathrm{coll}$ is the collision integral. Using the detailed balance condition, Eq.~(\ref{eq:BE}) is written as
\begin{equation}
e\bm E \cdot \bm v  \frac{ \partial f}{ \partial \varepsilon_{\bm k}}
=
    \int
\frac{d^2k^\prime}{(2\pi)^2}
W_{\bm k \bm k^\prime}
[f(\bm k) - f(\bm k')]
\label{eq:detailed_balance}
\end{equation}
We assume that the applied electric field is sufficiently weak so that the deviation of the distribution from equilibrium is small ($\delta f=f-f^0 \ll 1$). Then the collision integral can be approximated, within the generalized relaxation time approximation, as $\left(\frac{\partial f}{\partial t}\right)_\mathrm{coll}=-\frac{f-f^0}{\tau^i_{\varepsilon_{\bm k}} }$ where $\tau^i_{\varepsilon_{\bm k}}$ is the relaxation time along the $i$th direction with $i=H$ and $i=L$ denoting the heavy- and the light-mass directions, respectively. Using this, we find the solution of Eq.~(\ref{eq:BE}) to be in the form of
\begin{equation}
    f(\bm k) = f^0(\bm k) - e\tau^i_{\varepsilon_{\bm k}}\bm E \cdot \bm v  \frac{ \partial f}{ \partial \varepsilon_{\bm k}}
    \label{eq:distribution}
\end{equation}
By inserting Eq.~(\ref{eq:distribution}) into Eq.~(\ref{eq:detailed_balance}), we obtain the following integral equation for the relaxation time that fully takes into account the anisotropy of the system:

\begin{equation}
\frac{1}{\tau^i_{\varepsilon_{\bm k} } }
=
\int
\frac{d^2k^\prime}{(2\pi)^2}
W_{\bm k \bm k^\prime}
\left(
1 -
\frac{v^i_{\bm k^\prime} }{v^i_{\bm k^\prime} }
\frac{\tau^i_{\varepsilon_{\bm k^\prime} }}{\tau^i_{\varepsilon_{\bm k} } }
\right)
\label{eq:anisotropic_relaxation_time}
\end{equation}
where $v^i_{\bm k} = \partial\varepsilon_{\bm k}/\hbar\partial k_i$ is the velocity. Equation.~(\ref{eq:anisotropic_relaxation_time}) is the generalization of Eqs.~(\ref{eq:isotropic_relaxation_time}) and (\ref{eq:matrix_element}) from the isotropic Drude transport to the anisotropic case.
For a single band system, the current density is given by
\begin{equation}
    \bm j = -e\int \frac{d \bm k}{(2\pi)^2} \bm v(\bm k) f(\bm k)
    \label{eq:current_density}
\end{equation}
By substituting Eq.~(\ref{eq:distribution}) into Eq.~(\ref{eq:current_density}), it is easy to see that the dc conductivity at zero temperature is obtaind to be
\begin{equation}
\sigma_{ii}=
e^2\int \frac{d\bm k}{(2\pi)^2} 
\tau^i_{\varepsilon_{\bm k} }
v^i_{\bm k} v^i_{\bm k} 
\delta(\varepsilon_{\bm k} - \varepsilon_\mathrm{F})
\label{eq:conductivity}
\end{equation}
Since only the relaxation time evaluated at the Fermi surface contributes to the conductivity, in the following we present results only for $\tau^i_{\varepsilon_\mathrm{F}} (\theta_{k_\mathrm{F}})$
where $\theta_{k_\mathrm{F}}$ is the angle between the Fermi wavevector and the heavy-mass direction ($x$-axis). For a constant isotropic effective mass, Eq.~(\ref{eq:conductivity}) immediately gives the well-known Drude formula for the dc conductivity in terms of the transport scattering time: $\sigma=ne^2\tau/m$, where $n$ is the carrier density \cite{Ashcroft1976}.
\begin{figure*}[htb]
  \centering
  \includegraphics[width=0.7\linewidth]{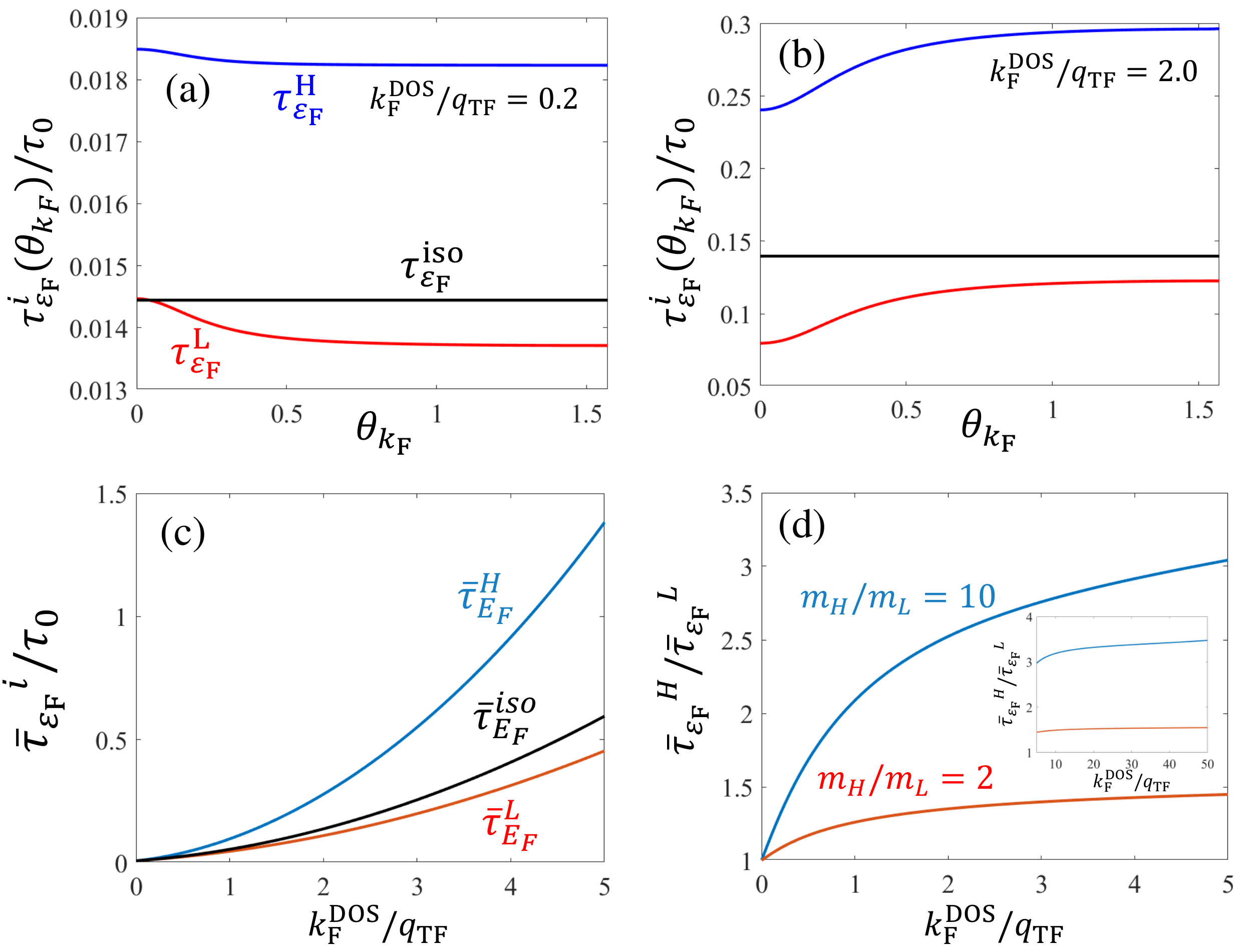}
  \caption{(a), (b) Plots of relaxation time as a function of $\theta$ for different values of $k_\mathrm{F}^\mathrm{DOS}$, i.e., electron densities. Here we set $m_\mathrm{H}/m_\mathrm{L}=10$ and the relaxation time is normalized by $\tau_0=\left[\frac{2\pi}{\hbar} n_\mathrm{imp}V_0^2 D(\varepsilon_\mathrm{F})\right]^{-1}$. (c) Plot of the angular average of the relaxation time as a function of $k_\mathrm{F}^\mathrm{DOS}$ for $m_\mathrm{H}/m_\mathrm{L}=10$, and (d) the ratio between the averaged relaxation times along the heavy-mass and the light-mass directions for $m_\mathrm{H}/m_\mathrm{L}=2$ (red) and $m_\mathrm{H}/m_\mathrm{L}=10$ (blue).}
  \label{fig:relaxation_time_theta}
\end{figure*}
\subsection{Short-range disorder}
We first consider $s$-wave short-range scatterers such as dislocation, point defect, atomic vacancy, etc. The impurity potential for short-range scatterers is extremely localized (actually zero range) in the real space, and thus in the momentum space it is given by a constant, i.e., $\left|V_{\bm k \bm k^\prime}\right|=V_0$. Then the relaxation time in Eq.~(\ref{eq:anisotropic_relaxation_time}) is expressed as
\begin{align}
\frac{1}{\tau_{\varepsilon_\mathrm{F}} (\theta_{k_\mathrm{F}}) }
=&
\frac{2\pi}{\hbar}
n_\mathrm{imp}
\int
\frac{d^2k^\prime}{(2\pi)^2}V_0^2 
\delta(\varepsilon_\mathrm{F} - \varepsilon_{\bm k^\prime} ) \notag\\
&\times \left(  1  -  
\frac{v^i_{\bm k^\prime}}{v^i_{\bm k}}
\frac{ \tau_{\varepsilon_\mathrm{F}} (\theta^\prime_{k_\mathrm{F}}) }
     	{  \tau_{\varepsilon_\mathrm{F}} (\theta_{k_\mathrm{F}}) }
\right)
\end{align}
Note that due to the symmetry of the system, the second term in the parenthesis is canceled out by integration, leading the relaxation time to be isotropic given by
\begin{align}
\frac{1}{\tau_{\varepsilon_\mathrm{F}}  }
=& 
\frac{2\pi}{\hbar}
n_\mathrm{imp}
\int
\frac{d^2k^\prime}{(2\pi)^2}V_0^2 
\delta(\varepsilon_\mathrm{F} - \varepsilon_{\bm k^\prime} ) \notag\\
=& 
\frac{2\pi}{\hbar}
n_\mathrm{imp}
\frac{V_0^2 D(E_\mathrm{F})}{2} \notag\\
=&\frac{2\pi}{\hbar}
n_\mathrm{imp}
\frac{V_0^2 m_\mathrm{DOS}}{2\pi\hbar^2}
\label{eq:short_relaxation_time}
\end{align}
This result shows that the isotropic approximation replacing anisotropic mass with the density-of-states mass works perfectly well for short-range disorder for all electron densities. Using Eq.~(\ref{eq:conductivity}) and Eq.~(\ref{eq:short_relaxation_time}), we can obtain the equation for the anisotropic short-range disorder resistivity
\begin{equation}
\rho_{ii}= 
\frac{2\pi n_\mathrm{imp}}{\hbar} 
\frac{V_0^2}{e^2 \varepsilon_\mathrm{F}}
m_i,
\label{eq:short_resistivity}
\end{equation}
which is the same formula as the isotropic resistivity with the effective mass being $m_i$ for $i$th direction. Note that the anisotropy of the resistivity ratio is the same as the mass ratio for all range of densities, i.e.,
\begin{equation}
\frac{\rho_\mathrm{H}}{\rho_\mathrm{L}}=\frac{m_\mathrm{H}}{m_\mathrm{L}}.
\label{eq:resistivity_ratio_short}
\end{equation}
We comment that the disappearance of the ($1 - \cos\theta$) vertex correction factor in the relaxation time here is the general result of $s$-wave scattering where the vertex correction vanishes. The direct prediction of our theory is that the short-range defect scattering induced anisotropic resistivity would reflect the precise mass anisotropy inserted into the isotropic  Drude formula with the resistivity in the heavy (light) mass direction being higher (lower) by the effective mass ratio.  It is easy to see that the resistivity in an arbitrary direction at an angle $\theta$ to the heavy mass direction (i.e., $x$-axis) will be given by the same standard Drude formula [Eq.~(\ref{eq:short_resistivity})] except that the effective mass will be $\theta$-dependent with $m(\theta)$ replacing $m_i$ in Eq.~(\ref{eq:short_resistivity}), where $1/m(\theta) = \cos^2\theta/m_\mathrm{H} + \sin^2\theta/m_\mathrm{L}$.


\subsection{Long-range disorder}
In this section we consider long-range charged Coulomb scatterers which often dominate the resistivity of 2D electronic materials \cite{Sarma2011}. We model the charged impurity potential as the RPA statically screened Coulomb potential given by $\left|V_{\bm k \bm k^\prime}\right|=\frac{2\pi e^2}{\varepsilon(\bm q)q}$ where $\bm q=\bm k - \bm k^\prime$ is the scattering wave vector. Then the anisotropic direction-dependent relaxation time [Eq.~(\ref{eq:anisotropic_relaxation_time})] is written as
\begin{align}
\frac{1}{\tau_{\varepsilon_\mathrm{F}} (\theta_{k_\mathrm{F}}) }
=&
\frac{2\pi}{\hbar}
n_\mathrm{imp}
\int
\frac{d^2k^\prime}{(2\pi)^2}
\left[ \frac{2\pi e^2}{\varepsilon(\bm q)q}\right]^2 
\delta(\varepsilon_\mathrm{F} - \varepsilon_{\bm k^\prime} ) \notag\\
&\times \left[  1  -  
\frac{v^i_{\bm k^\prime}}{v^i_{\bm k}}
\frac{ \tau_{\varepsilon_\mathrm{F}} (\theta^\prime_{k_\mathrm{F}}) }
     	{  \tau_{\varepsilon_\mathrm{F}} (\theta_{k_\mathrm{F}}) }
\right].
\label{eq:relaxation_time_longrange}
\end{align}
The second term in the parenthesis in Eq.~(\ref{eq:relaxation_time_longrange}) is not canceled out by integration unlike the case for short-range disorder because the vertex correction does not vanish for the long-range non-$s$-wave scattering. The easiest way to solve Eq.~(\ref{eq:relaxation_time_longrange}) is to discretize the angular variable $\theta_{k_\mathrm{F}}$ so that Eq.~(\ref{eq:relaxation_time_longrange}) is converted into a standard discretized matrix eigenvalue problem \cite{Vyborny2009, Park2017}.

Figures~\ref{fig:relaxation_time_theta}(a) and \ref{fig:relaxation_time_theta}(b) show the numerically calculated relaxation time as a function of $\theta_{k_\mathrm{F}}$ for $k_\mathrm{F}^\mathrm{DOS}/q_\mathrm{TF}=0.2$ and $2.0$. Note that in contrast to the short-range relaxation time [Eq.~(\ref{eq:short_relaxation_time})], the long-range relaxation time varies rapidly with $\theta_{k_\mathrm{F}}$ exhibiting a manifest anisotropy arising from mass anisotropy. This angular-dependent behavior of the relaxation time cannot be reproduced by the isotropic approximation (black line). When the electron density is low [Fig~\ref{fig:relaxation_time_theta}(a)], the relaxation time is maximum at $\theta=0$, monotonically decreasing with increasing $\theta$, and being the minimum at $\theta=\pi/2$. At a higher density [Fig~\ref{fig:relaxation_time_theta}(b)] the relaxation time behaves in the opposite manner, being smallest at $\theta=0$ and maximum at $\theta=\pi/2$. One can also see by comparing Fig~\ref{fig:relaxation_time_theta}(a) with Fig~\ref{fig:relaxation_time_theta}(b) that the relaxation time varies more rapidly as a function of $\theta_{k_\mathrm{F}}$ at a higher density, implying that the anisotropy becomes stronger as the density increases. This can be understood by looking at the screened Coulomb potential in Eq.~(\ref{eq:relaxation_time_longrange}). Using the Thomas-Fermi approximation, the screened Coulomb potential is written as $v_\mathrm{sc}(\bm q)=\frac{2\pi e^2}{q + q_\mathrm{TF}}$. Note here that the screened potential in the Thomas-Fermi limit is isotropic since the static dielectric function is isotropic at small momenta as discussed in Sec.~\ref{sec:static_screening}.
For sufficiently low densities where $q_\mathrm{TF}\gg k_\mathrm{F}^\mathrm{DOS} $, the screened Coulomb potential can be further approximated as a constant
\begin{equation}
v_\mathrm{sc}(\bm q)=\frac{2\pi e^2}{q + q_\mathrm{TF}}\approx\frac{2\pi e^2}{q_\mathrm{TF}},
\label{eq:screened_potential}
\end{equation}
which is so called ``complete screening approximation''. This shows that in the low density limit the 2D impurity potential for long-range Coulomb disorder is given by a constant in the momentum space similar to the short-range impurity potential. The short-range nature of the screened Coulomb disorder in the low-density limit appears counter-intuitive, but is true by virtue of the fact that $k_\mathrm{F}$ in 2D goes as the square root of carrier density but $q_\mathrm{TF}$ is a constant.
By substituting $V_0$ in Eq.~(\ref{eq:short_relaxation_time}) with Eq.~(\ref{eq:screened_potential}) we obtain the long-range relaxation time in the low density limit
\begin{align}
\frac{1}{\tau_{\varepsilon_\mathrm{F}}  }
=\frac{2\pi}{\hbar}
n_\mathrm{imp}
\frac{\pi \hbar^2}{2m_\mathrm{DOS}},
\label{eq:long_relaxation_time_in_strong}
\end{align}
which is isotropic and constant, similar to the short-range relaxation time. This indicates that the isotropic approximation using the isotropic density-of-state mass works well for long-range disorders when the electron density is sufficiently low. Note that such an isotropic low-density strongly screened behavior of the long-range relaxation time comes from the 2D screening becoming isotropic in the long-wavelength limit which is discussed in-depth in Sec.~\ref{sec:static_screening}.


\begin{figure}[htb]
  \centering
  \includegraphics[width=0.95\linewidth]{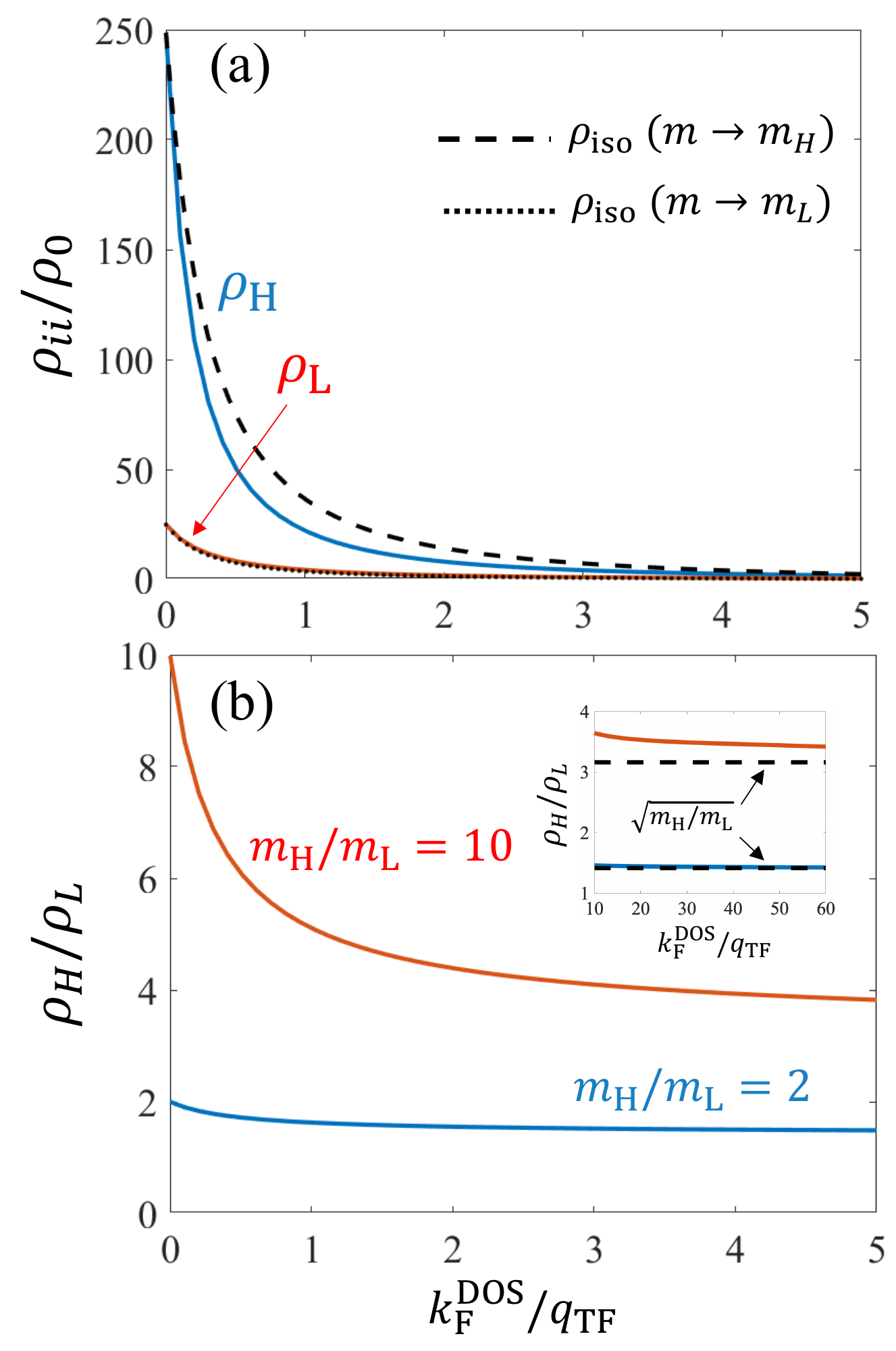}
  \caption{(a) Calculated resistivity along the heavy-mass (blue) and the light-mass (red) directions along with the isotropic resistivity (black) calculated with the effective isotropic mass being $m_\mathrm{H}$ and $m_\mathrm{L}$ for $\rho_\mathrm{H}$ and $\rho_\mathrm{L}$, respectively. Here we set $m_\mathrm{H}/m_\mathrm{L}=10$, and the resistivity is normalized by $\rho_0=\frac{\pi \hbar n_\mathrm{imp}}{2e^2k_\mathrm{F}^2}$ (b) The anisotropy ratio of the resistivity for $m_\mathrm{H}/m_\mathrm{L}=2$ (blue) and $m_\mathrm{H}/m_\mathrm{L}=10$ (red). The inset in (b) shows the the convergence of the resistivity ratio to the square root of the mass ratio in the high density limit (i.e.,$k_\mathrm{F}^\mathrm{DOS}\gg q_\mathrm{TF}$). }\label{fig:resistivity}
\end{figure}

Figure~\ref{fig:relaxation_time_theta}(c) shows the relaxation time averaged over $\theta_{k_\mathrm{F}}$, i.e.,
\begin{equation}
\overline{\tau}^i_{\varepsilon_\mathrm{F}}
=
\frac{1}{2\pi}
\int^{2\pi}_0 d\theta 
\tau^i_{\varepsilon_F}(\theta_{\varepsilon_\mathrm{F}})
\end{equation}
along the heavy-mass and the light-mass directions as a function of $k_\mathrm{F}^\mathrm{DOS}$. 
Note that $\overline{\tau}^\mathrm{H}_{\varepsilon_\mathrm{F}}> \overline{\tau}^\mathrm{L}_{\varepsilon_\mathrm{F}}$ always with $\tau_{\varepsilon_\mathrm{F}}^\mathrm{iso}$ lying in between them. The relaxation time along the heavy-mass direction increases more rapidly than that along the light-mass direction with increasing electron density. This can be more explicitly seen in Fig.~\ref{fig:relaxation_time_theta}(d) where we plot the anisotropy ratio of the relaxation time. In a reasonable range of density up to $k_\mathrm{F}^\mathrm{DOS}=50q_\mathrm{TF}$, the anisotropy of the relaxation time is smaller than the mass anisotropy, i.e., $\overline{\tau}^\mathrm{H}_{\varepsilon_\mathrm{F}}/\overline{\tau}^\mathrm{L}_{\varepsilon_\mathrm{F}}<m_\mathrm{H}/m_\mathrm{L}$ [see the inset in Fig.~\ref{fig:relaxation_time_theta}(d)]. Note that the relaxation time becomes isotropic, as expected, in the zero-density strong-screening limit with $\overline{\tau}^\mathrm{H}_{\varepsilon_\mathrm{F}}/\overline{\tau}^\mathrm{L}_{\varepsilon_\mathrm{F}}$ converging to unity as $k_\mathrm{F}^\mathrm{DOS}/q_\mathrm{TF}\rightarrow0$.

Figure.~\ref{fig:resistivity}(a) shows the calculated resistivity  as a function of $k_\mathrm{F}^\mathrm{DOS}$. All the curves show the typical behavior of the resistivity, diverging at zero density and decreasing monotonically with increasing density. We find that $\rho_\mathrm{H}>\rho_\mathrm{L}$ always for all density ranges in spite of the relaxation time itself satisfying $\overline{\tau}^\mathrm{H}_{\varepsilon_\mathrm{F}}> \overline{\tau}^\mathrm{L}_{\varepsilon_\mathrm{F}}$. This is because the resistivity goes as $m/\tau$, and the relaxation time being larger does not necessarily imply a lower resistivity because of the mass factor. The black dashed and dotted lines show the resistivity calculated using the existing isotropic results with the effective mass given by $m_\mathrm{H}$ and $m_\mathrm{L}$ for $\rho_\mathrm{H}$ and $\rho_\mathrm{L}$, respectively. Note that the isotropic resistivity results well reproduce the anisotropic resistivity in the low density regime where the long-range Coulomb disorder is strongly screened whereas it fails when the density becomes higher, especially along the heavy-mass direction.
In Fig.~\ref{fig:resistivity}(b) we present the calculated long-range resistivity ratio for two different mass ratios $m_\mathrm{H}/m_\mathrm{L}=2$ and $10$. In the zero density limit, the anisotropy of the long-range disorder limited resistivity is equal to the mass ratio as that of the short-range resistivity, which can be easily proven by substituting $V_0$ in Eq.~(\ref{eq:short_resistivity}) with the strongly screened Coulomb potential Eq.~(\ref{eq:screened_potential}). 
Note that the anisotropy of the resistivity is suppressed with increasing density. The amount of suppression is larger for larger mass ratios. In the inset of Fig.~\ref{fig:resistivity}(b), we plot the resistivity ratio up to very high density $k_\mathrm{F}^\mathrm{DOS}=60q_\mathrm{TF}$ showing that $\rho_\mathrm{H}>\rho_\mathrm{L}$ saturates to $\sqrt{m_\mathrm{H}/m_\mathrm{L}}$ as the density increases. This result indicates that the anisotropy ratio for long-range scattering limited resistivity is between $m_\mathrm{H}/m_\mathrm{L}$ and $\sqrt{m_\mathrm{H}/m_\mathrm{L}}$ in the low and high density limits. 
Note that the anisotropy ratio of the long-range resistivity strongly depends on the electron density in contrast to that of the short-range resistivity, which is just given by the mass anisotropy ratio regardless of the electron density. Using such contrasting resistivity behaviors, one may possibly distinguish between short-range and long-range disorder in the sample using a measurement of transport as a function of density. In a recent experiment on GeAs, which exhibits a strong anisotropy, it has been reported that the resistivity anisotropy ratio varies as a function of electron density \cite{Sun2020}. Our results imply that the long-range disorder-limited resistivity in the sample contributes to such a density dependent behavior of the resistivity anisotropy ratio.

The square root mass ratio dependence of the high-density long-range disorder-limited resistivity can be analytically understood as arising from the fact that for high densities screening weakens by virtue of $k_\mathrm{F}\gg q_\mathrm{TF}$, and therefore, the disorder potential is effectively unscreened.  Such an unscreened Coulomb disorder leads to a 2D resistivity going as $\rho \sim m^{1/2}$ simply on dimensional ground as can be shown easily by following in arguments in Ref.~\cite{Sarma2013}. This leads to the very high density asymptotic resistivity ratio approach $(m_\mathrm{H}/m_\mathrm{L})^{1/2}$ as we find numerically.  Given that the limiting resistivity in the low- and high-density limit goes as $m_\mathrm{H}/m_\mathrm{L}$ and $(m_\mathrm{H}/m_\mathrm{L})^{1/2}$ respectively, the actual resistivity ratio at any physical density will be less than the effective mass anisotropy, i.e., $m_\mathrm{H}/m_\mathrm{L} > \rho_\mathrm{H}/ \rho_\mathrm{L} > (m_\mathrm{H}/m_\mathrm{L})^{1/2}$, which provides a weak rationale for neglecting mass anisotropy in transport theories of 2D systems

Interestingly, and perhaps coincidentally, this square root transport behavior of the anisotropy we find in the unscreened high-density limit at zero magnetic field, has also been found in both experiments and theories of the high-field  composite fermions in 2D anisotropic systems at half-filled Landau levels \cite{Yang2013, *Jo2017, *Ippoliti2017}. Whether this agreement between our Boltzmann transport theoretic for unscreened Coulomb scattering and the composite fermion results in high magnetic field is a coincidence or a deep connection is unknown and worthy of future theoretical consideration.  It is worth noting in this context that our finding of the transport anisotropy going as the effective mass anisotropy in the completely screened low-density limit also agrees with certain Chern-Simons theories for composite fermions at half-filling \cite{Balagurov2000}.

\section{Summary and Conclusion} \label{sec:conclusion}
In this paper, we have provided theoretical studies of screening properties, RKKY interaction, and Drude transport properties of an anisotropic 2D electron gas as occurring in 2D metals and doped semiconductors.. 

We first investigated the screening properties by obtaining the exact form of the static polarizability and the corresponding RPA dielectric function. In the long-wavelength limit, the polarizability exhibits a constant isotropic behavior, leading the screening to be isotropic despite the anisotropy of the system. The polarizability loses its isotropy above a critical momentum $q_c(\theta)$ where the polarizability suddenly drops. The screening is stronger along the heavy-mass direction than along the light-mass direction. The resultant Friedel oscillations of the screened potential manifest strongly anisotropic oscillations with dramatic angle-dependent spatial periodicity, which should have direct experimental consequences.

We also obtained the RKKY interaction which is the indirect exchange interaction between localized magnetic moments mediated by the itinerant electrons. The key anisotropic features are i) the period of the RKKY oscillation along the heavy- and light-mass directions is determined by the magnitude of the Fermi wavevector along the corresponding direction, (i.e., $\pi/k^\mathrm{i}_\mathrm{F}$ along the $i$th direction), and ii) the RKKY interaction decays much faster along the heavy-mass direction with its decay rate given by $\sim1/(2k_\mathrm{F}^ir)^2$ along $i$th direction. This should have experimentally observable consequences in 2D anisotropic materials.

We then presented the theory for transport properties of anisotropic 2DEG using the Boltzmann transport theory fully incorporating the anisotropic mass. We consider two types of scatterers: short-range disorder and long-range Coulomb disorder. We find that the short-range relaxation time at the Fermi surface is perfectly isotropic, being actually the same as the existing result for an isotropic 2DEG with the effective mass replaced by the density-of-states mass. Therefore, the short-range resistivity is proportional to the mass corresponding to the direction of the applied electric field, i.e., $\rho\sim m_i$ and thus the resistivity ratio is always the same as the mass ratio (i.e., $\rho_\mathrm{H}/\rho_\mathrm{L}=m_\mathrm{H}/m_\mathrm{L}$).
For the long-range scatterers, we find that the relaxation time is strongly anisotropic, varying as a function of the momentum direction on the Fermi surface ($\theta_{k_\mathrm{F}}$). We find $\tau_\mathrm{H} > \tau_\mathrm{L}$ always, and the anisotropy of the relaxation time is reduced as the density decreases (i.e., as the screening becomes stronger). We find that the long-range resistivity ratio varies with electron density: in the zero density limit the resistivity ratio is the same as the mass ratio (i.e., $\rho_\mathrm{H}/\rho_\mathrm{L}=m_\mathrm{H}/m_\mathrm{L}$) whereas it is equal to the square root of the mass ratio (i.e, $\rho_\mathrm{H}/\rho_\mathrm{L}=\sqrt{m_\mathrm{H}/m_\mathrm{L}}$) in the high density limit. In the intermediate density regime, the resistivity ratio lies in between these two values. Our results show that the resistivity for the short-range and long-range disorder has quite different dependence on electron density, and thus one may be able to distinguish between the two scatterers through measuring the transport as a function of density. In general, $\rho_\mathrm{H} > \rho_\mathrm{L}$ always although at the same time $\tau_\mathrm{H} > \tau_\mathrm{L}$ also always.  But, generically, at any arbitrary density the resistivity ratio, with $m_\mathrm{H}/m_\mathrm{L}> \rho_\mathrm{H}/\rho_\mathrm{L}>(m_\mathrm{H}/m_\mathrm{L})^{1/2}$, is suppressed compared with the effective mass anisotropy, offering a weak justification of the widespread use of the isotropic transport approximation in the theoretical literature.


\section{Acknowledgments} \label{sec:acknowledgement}
This work is supported by the Laboratory for Physical Sciences. The authors thank Ted Einstein for asking about the anisotropic RKKY interactions.

\bibliographystyle{apsrev4-2}
\bibliography{ref}

\end{document}